\documentclass[manuscript]{acmart}
\usepackage{hyperref}

\AtBeginDocument{%
  }

\setcopyright{none}
\acmDOI{}
\acmISBN{}

\acmConference[XAIxArts 2026]{Explainable AI for the Arts Workshop 2026}{July 13, 2026}{London, UK}

\begin{document}

\title[Who Chooses the Artwork?]{Who Chooses the Artwork? Curatorial Agency and Distributed Intent in Botto}

\author{Haoting Alexa Yu}
\affiliation{%
  \institution{Creative Computing Institute, University of the Arts London}
  \city{London}
  \country{UK}
}
\email{h.yu0620251@arts.ac.uk}

\author{Bea Wohl}
\affiliation{%
  \institution{Creative Computing Institute, University of the Arts London}
  \city{London}
  \country{UK}
}
\email{b.wohl@arts.ac.uk}

\renewcommand{\shortauthors}{Yu and Wohl}

\begin{abstract}
Botto is often described as a decentralized autonomous artist, but its authorship cannot be located in image generation alone. This paper examines Botto as an agentic curatorial system in which generation, ranking, voting, feedback, and minting form a recursive loop. Drawing on Botto’s documentation and prior accounts of the project, we argue that agency in Botto is distributed but asymmetrical: community participants influence artistic direction through voting and governance, while Botto’s internal models structure which images become visible, selectable, and recognizable as works. The case shows that, for AI art systems, explainability should include not only model behavior but also the curatorial procedures through which outputs acquire artistic status.
\end{abstract}


\maketitle

\begin{figure}[H]
  \centering
  \includegraphics[width=\linewidth]{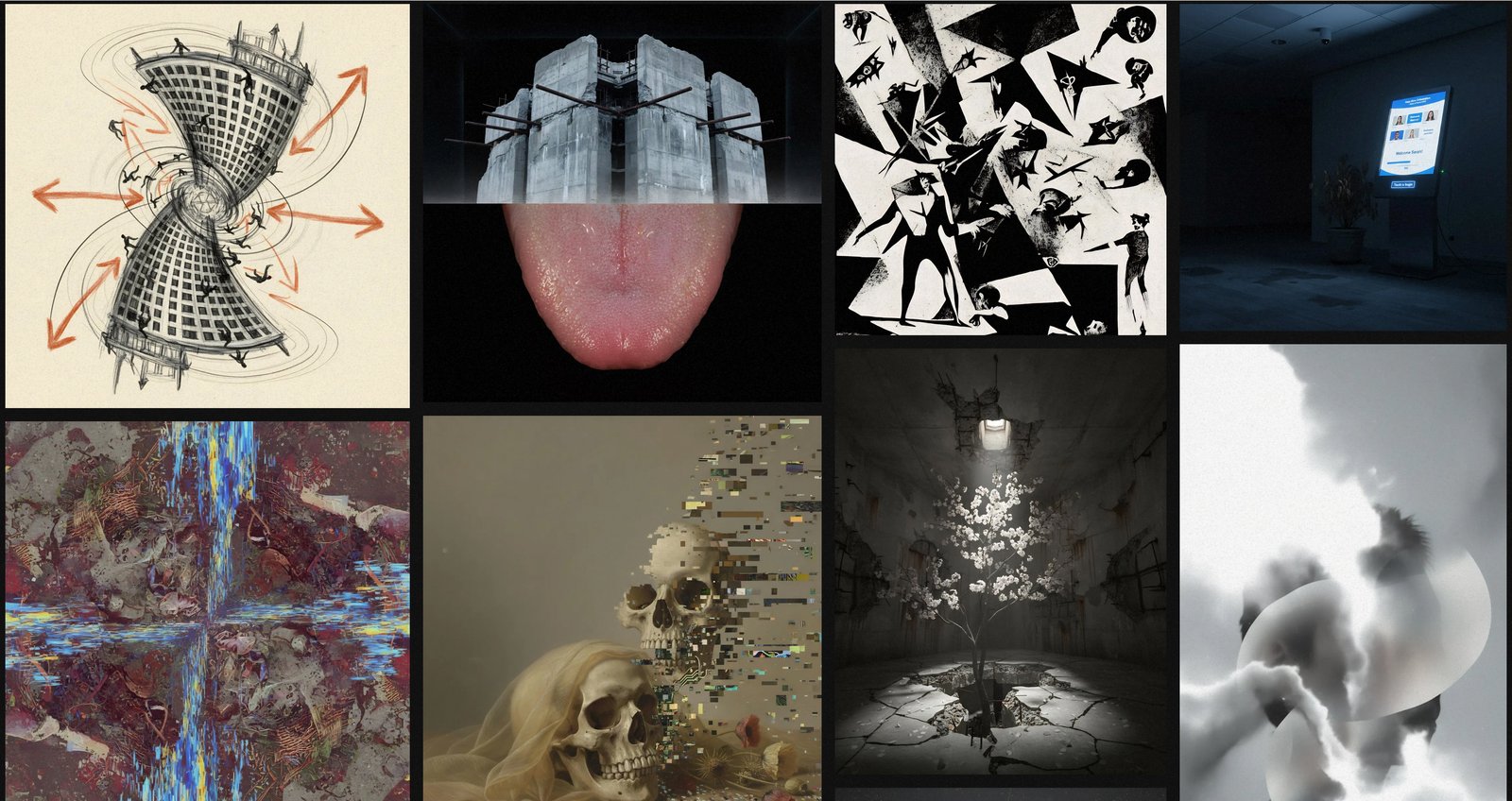}
  \caption{A selection of Botto's artworks, illustrating the stylistic range the system produces across its Periods. Image \textcopyright{} BottoDAO; reproduced for scholarly analysis.}
  \Description{A grid of eight artworks by Botto in widely varying styles, including a pencil-sketch architectural form, a concrete structure above a tongue, a black-and-white angular composition, a dim gallery installation photograph, an abstract colour field, a glitching skull, a blossoming tree in a cracked room, and a soft white abstract form.}
  \label{fig:teaser}
\end{figure}

\section{Introduction: Botto as an Agentic Curatorial System}

\begin{figure}[t]
  \centering
  \includegraphics[width=\linewidth]{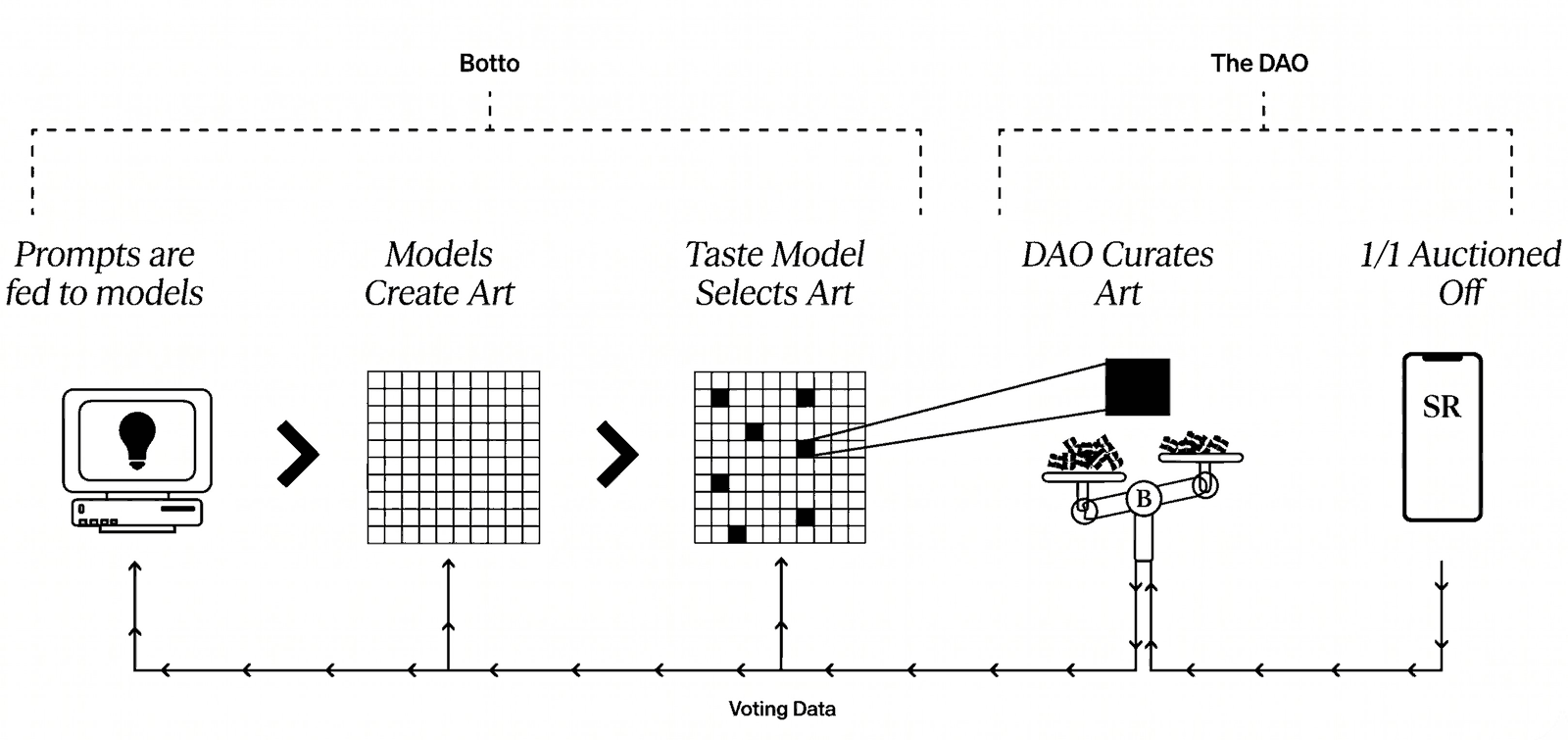}
  \caption{Botto's own infographic of its weekly production cycle~\cite{botto2024botto101}. From left to right: prompts are fed to text-to-image models; the models generate a large pool of candidate images; the taste model pre-selects a shortlist (the black cells); the DAO votes to curate a single work; and the winning piece is auctioned on SuperRare. The ``Voting Data'' arrow along the bottom indicates that votes feed back into earlier stages. The diagram shows the standard weekly cycle; it does not depict the thirteenth-week self-curation discussed in Sections 2 and 3. Image \textcopyright{} BottoDAO; reproduced for scholarly analysis.}
  \Description{A horizontal five-stage flow diagram. A computer feeds prompts to models; a grid represents a large pool of generated images; the same grid with a few highlighted cells represents the taste model's shortlist, one cell enlarged; a balance scale labelled B represents the DAO voting to curate; and a phone labelled SR represents the work auctioned on SuperRare. A horizontal arrow labelled Voting Data runs along the bottom back to the earlier stages.}
  \label{fig:diagram}
\end{figure}

Botto operates as a decentralized image-making system governed by a community of participants who review and vote on its outputs. Each week, it generates a large pool of candidate images. Botto's taste model trained on previous feedback selects 350 of these to enter the voting pool for community review and voting. The resulting vote determines which image is formally recognized and minted as that week's work and shapes subsequent rounds of prompt generation and image ranking (Figure~\ref{fig:diagram}) \cite{botto2025artengine,klingemann2022botto}. Botto's production is organized into named Periods, each defined by particular themes, visual styles, and image generation methods (Figure~\ref{fig:teaser}) \cite{botto2026periods}. Most weeks follow the same cycle of generation, ranking, and community voting. In recent Periods, the final week modifies this arrangement: community voting is suspended, and Botto's taste model selects the final work from the system's own prior outputs, choosing the image with the strongest cumulative relationship to the Period's twelve preceding works \cite{botto2026voting,botto2024interstice,botto2024synthetic}.

This architecture means that Botto is better understood as an agentic curatorial system than as a one-shot image generator. Generation, ranking, voting, feedback, and recognition are tightly coupled. AI models participate in both production and selection, while audiences participate in both evaluation and the shaping of future outputs. As these functions become entangled, the boundary between creator and audience becomes less clear, but the model and the audience do not shape the work in the same way or to the same degree. Recent XAIxArts scholarship has moved beyond narrowly technocentric notions of explanation. It approaches explainability through artistic practice, asking how artists work with AI systems and how understanding takes shape in that process \cite{abuzuraiq2025,broad2024,bryankinns2025,zhanglong2025}. This shift is important for Botto because the system operates across multiple stages, incorporates feedback, and changes its own trajectory over time. This paper asks where authorial agency is exercised when a work emerges through this recursive process. In Botto, authorship is exercised across the curatorial loop that filters outputs, selects one, and recognizes it as the work. The key issue is how one image, out of many, comes to count as the work.

\section{Distributed but Asymmetrical Agency}
The claim that Botto is an agentic curatorial system becomes clearer once we distinguish between distributed agency and equal agency. Botto is governed by BottoDAO, the project's token-based governance community, and that community does more than respond to finished images. The DAO participates in Period-level direction, while weekly art voting feeds back into both prompt construction and the taste model that pre-selects images for review \cite{botto2025artengine,botto2026periods}. Therefore, authorship in Botto is recursive: community judgment is folded into later production rather than remaining outside it. This is consistent with Klingemann, Hudson, and Epstein's early description of Botto as a decentralized autonomous artist whose generative, evaluative, and governance functions are closely linked \cite{klingemann2022botto}.

Additionally, the community only sees a small part of what Botto produces. Current documentation states that Botto can generate as much as 70{,}000 images in a week and that the taste model pre-selects 350 images for community voting; the earlier 2022 account describes the same structure at a smaller scale, with roughly 4{,}000 weekly images filtered to 350 finalists \cite{botto2026periods,botto2025artengine,klingemann2022botto}. The change in scale is important because the field of generation has expanded, while the number of newly presented images for weekly public judgment remains limited. The community is therefore responding to a pre-selected set whose visibility has already been shaped by the system. Human participants still take part in the process, but their judgment is strongly conditioned by the model's prior filtering and ranking.

The community's influence is divided across two mechanisms. The first is weekly art voting, which helps decide which candidate image becomes the weekly artwork. In this process, participants use Voting Points (VP) to vote on the fragments selected for community review. VP can be understood as internal voting credits in the Botto app. They are generated through participation in the Botto ecosystem, including staking \$BOTTO, providing BOTTO--ETH liquidity, holding an Access Pass or Pipe, or logging into the Botto app \cite{botto2026vp}. The second mechanism is formal governance. For project-level decisions, BottoDAO uses Snapshot votes. In Botto's Snapshot votes, eligibility depends on active staking, and voting weight scales with the amount of \$BOTTO represented by a participant's stake \cite{botto2025snapshot}. These two mechanisms give the community influence in different ways. Weekly art voting shapes which image is selected and minted as the next artwork, while governance voting shapes the broader rules and direction of the system. The two mechanisms also differ in their transparency: Snapshot governance votes are publicly recorded and verifiable, whereas VP-based weekly art voting takes place within the Botto app and is not exposed for external inspection.

From Botto's model training pipeline, the documentation states that weekly votes are used to train the taste model, but training influence is not only proportional to VP weight. It is designed to avoid excessive influence from a single voter, which means many separate voters can carry more training significance than a larger number of votes from one account \cite{botto2025artengine}. Botto also includes unexpected outputs so that the system does not narrow too quickly into one dominant taste profile \cite{botto2025artengine}. Recent additions such as the Knowledge Graph and Creative Reasoning framework further give Botto structures for memory, prompt refinement, and iterative development \cite{botto2025artengine,botto2025creativereasoning}. For that reason, authorship in Botto is best understood as distributed but asymmetrical: human participants influence the system through voting and governance, while Botto's internal architecture shapes visibility, selection, and aesthetic possibility.

\section{Selection, Recognition, and the Status of the Work}
This asymmetry becomes clearest when the question shifts from who generated the image to how one image comes to count as the work. Botto produces far more images than it ever recognizes. Most generated images are never presented to the community and never enter the project's canonical archive. Each week, one fragment is formally minted as the final artwork. Its title is generated algorithmically through random word combinations evaluated with CLIP (Contrastive Language--Image Pre-training), and its description is selected by the DAO from a small set of LLM-generated options that CLIP identifies as fitting the image, with minor human correction for typos and punctuation \cite{botto2025artengine}. Therefore, an image becomes the work through ranking, selection, naming, and formal recognition. Botto's authorship here lies in deciding where the boundary falls between a generated image and a recognized artwork.

This logic has become more explicit in recent changes to Botto's Period structure. Botto's voting documentation states that, on the thirteenth week, Botto selects its own work to mint and community voting pauses \cite{botto2026voting}. Period recaps explain this process in more detail: beginning with the Interstice Period, Botto self-curates the thirteenth and final mint by analyzing the millions of images it has generated since its origin and using its taste model to select the work with the strongest cumulative relationship to the twelve previous mints of that Period \cite{botto2024interstice,botto2024synthetic}. This does not provide a full technical account of the taste model, since the exact scoring procedure is not fully disclosed. However, it makes the curatorial rule visible: the final mint is selected not through community preference in that week, but through Botto's retrospective evaluation of relation and fit within the Period.
The thirteenth mint also foregrounds a question central to recent XAIxArts scholarship. Bryan-Kinns et al. argue that explainability in the arts should move beyond technocentric and mechanistic accounts \cite{bryankinns2025}. Botto fits that shift well because explainability within the system is procedural and curatorial, which means the method used to decide what counts as the work becomes part of the artwork's meaning.

\section{Conclusion}
The paper has argued that authorship in Botto is distributed but asymmetrical. The DAO influences the system through weekly voting and governance decisions, yet Botto’s architecture also limits what the DAO encounters in the first place: only certain images are promoted into view, and only some of those can become works. For XAIxArts, the implication is that explainability in AI art cannot stop at model outputs or technical generation. It also has to explain curation: why the taste model advances a particular image into the voting pool, and why the thirteenth-week mint is judged the strongest fit for the Period, even where the exact scoring is not disclosed. This account draws on Botto's own documentation, and independent evidence such as analysis of voting patterns or interviews with DAO participants is a clear direction for future work.

\bibliographystyle{ACM-Reference-Format}
\bibliography{xaixarts2026-botto}

\end{document}